\documentclass{article}

\usepackage{amssymb}
\usepackage{amsfonts}
\usepackage{amsmath}
\usepackage{graphicx}
\usepackage{latexsym}

\title{Exact solutions of the \\
3-wave resonant interaction 
equation}

\author{Antonio Degasperis$^{(1)}$, Sara Lombardo$^{(2,3)}$\\
$^{(1)}$ Dipartimento di Fisica, Universit\`a di Roma 
``La Sapienza'', \\
and Istituto Nazionale di Fisica Nucleare, Sezione di 
Roma,\\ 
Rome, Italy.\\
E-mail: antonio.degasperis@roma1.infn.it\\ 
\\
$^{(2)}$ Department of Applied Mathematics, University of
Leeds, Leeds, UK, \\
$^{(3)}$ Department of Mathematics, Vrije Universiteit, 
Amsterdam, NL.\\ E-mail: sara@few.vu.nl}

\date{}

\begin{document}

\maketitle
    
\begin{abstract}
The Darboux--Dressing Transformations are applied to the Lax pair
associated to the system of nonlinear equations describing the 
resonant interaction of three waves in 1+1 dimensions. We display  explicit 
solutions featuring  localized waves whose profile vanishes 
at  the spacial boundary $|x|=\infty$, and which are \emph{not} pure soliton 
solutions. These solutions depend on an arbitrary function and allow to deal with collisions of waves with various profiles.
\end{abstract}



\section{Introduction}
The propagation in 1+1 dimensions of three resonating waves is 
modelled by
the following system of three coupled PDEs
\begin{equation}\label{3wri}
D_{n}\,\chi_{n}=g_{n}\,\chi _{n+1}^{\ast }\,\chi_{n+2}^{\ast}\, ,
\quad n=1,2,3\,\,\,\textrm{mod}\left( 3\right)\, ,  
\end{equation}
where the three dependent variables  $\chi_{n} \equiv \chi_{n}(x,t)$ 
are generally  
complex functions of two real independent variables, $x$, 
(\emph{space}) and $t$
(\emph{time}). The three differential operators $D_n$, $n=1,\,2,\,3$, 
in their linear (i.e. left-hand-side)
part are first--order and read
\begin{equation}\label{diffop}
D_{n}:=\partial_{t}+c_{n}\partial_{x},
\end{equation}
where the three
\emph{real} constants
$c_{n}$ are taken all different among themselves.  Each $c_{n}$ is 
the 
characteristic velocity 
at which the field $\chi_{n}$ propagates in the space--time region 
where no interaction takes place. Here and
hereafter a subscripted independent variable denotes partial
differentiation with respect to it, asterisks denote complex 
conjugation
and all indices are considered modulo $3$. The
three constants $g_{n}$ are generally complex and have the 
significance of
\emph{coupling constants}. Their values depend of course on the 
physical
context to which the 3Wave Resonant Interaction (3WRI) equation
applies. It is remarkable however that, if these coupling constants
satisfy certain conditions (see below), then the 3WRI equation
(\ref{3wri}) is integrable. In
this paper we consider only such case and apply Darboux (or 
``dressing")
transformations to construct explicit solutions of
the system (\ref{3wri}). 

The conditions on the complex constants $g_{n}$ which are necessary 
and
sufficient for the integrability of the 3WRI equation (\ref{3wri}) 
are:
\begin{equation}\label{intcond}
g_n=\sigma_n |g_n| \exp(i\Gamma)\,\,,\,\,\, \sigma_n^2=1\,\,,\,\,
n=1,2,3\, .
\end{equation}
Here the three signs $\sigma_n$ and the three moduli $|g_n|$ are
\emph{arbitrary} and the condition is that the (also \emph{arbitrary})
phase $\Gamma$ is common to the three $g_n$'s. It is worth pointing 
out
here that these conditions do single out only one integrable system in
the following sense: the structure of the system (\ref{3wri}) is left
unchanged by the transformation
\begin{equation}\label{trans}
\bar{\chi}_n(\bar{x},\bar{t})=\alpha_n \chi_n(x,t)\,\,,\,\,\,
\bar{x}=Ax+Bt\,\,,\,\, \bar{t}=Cx+Dt\,,
\end{equation}
where the three arbitrary constants $\alpha_n$ are complex and the 
four
constants $A,B,C,D$ are real and arbitrary except for the requirement
$AD-BC\neq 0$; indeed it is clear that such transformation only 
changes
the value of the six constants $c_n$ and $g_n$. It is easily seen that
the transformation (\ref{trans}), as it should, takes $g_n$'s 
satisfying
the integrability conditions (\ref{intcond}) into different $g_n$'s 
which
satisfy the same conditions, and that any set of values of $g_n$'s
satisfying (\ref{intcond}) can be transformed into any other set which
satisfy (\ref{intcond}), for instance $g_1=g_2=g_3=1$. Without any 
loss of
generality, hereafter we choose the coupling constants to have the 
expressions  
 \begin{equation}\label{gs}
g_{n}=c_{n+1}-c_{n+2}\,,\quad n=1,2,3\,\,\,\textrm{mod}\left( 
3\right)\,  
\end{equation}
in terms of the velocities $c_n$ with the implication that
$g_1+g_2+g_3=0$ and, therefore, that the $g_n$'s cannot have all the 
same
sign. The reason for this choice is practical as the Lax pair 
associated
to the system (\ref{3wri}) \cite{ZM} in this case takes a simple look and it 
reads
\begin{subequations}
\label{laxpair}
\begin{equation}\label{laxpaira}
\Psi_{x}=(-ikB+U(x,t))\Psi\, ,
\end{equation}
\begin{equation}\label{laxpairb}
\Psi_{t}=(ikA+V(x,t))\Psi\, ,
\end{equation} 
\end{subequations}
where $A$ and $B$ are two diagonal, real, traceless matrices
\begin{equation}\label{AB}
\begin{array}{c}
    A=\textrm{diag}\{a_{1},  a_{2}, a_{3}\}\,,\quad B=\textrm
{diag}\{b_{1}, b_{2}, b_{3}\}\,, \\
 \\
    \, A=A^{\ast},\quad B=B^{\ast}\,,\,\quad
\textrm{tr}(A)=\textrm{tr}(B)=0\, . 
\end{array}
\end{equation}
The $3\times3$ matrix $\Psi\equiv \Psi(x,t,k)$ is the common
solution of the two ODEs (\ref{laxpair}) while 
$U(x,t)$ and $V(x,t)$ are off-diagonal matrices whose entries are 
related
to the three wave fields $\chi_{n}(x,t)$:
\begin{equation}\label{UV}
U=\left( \begin{array}{ccc}
0 & \chi_{3} & -\chi_{2}^{*}\\
-\chi_{3}^{*} & 0 & \chi_{1}\\
\chi_{2} & -\chi_{1}^{*} & 0\end{array}\right) \, ,\quad  
V=\left( \begin{array}{ccc}
0 & -c_{3}\chi_{3} & c_{2}\chi_{2}^{*}\\
c_{3}\chi_{3}^{*} & 0 & -c_{1}\chi_{1}\\
-c_{2}\chi_{2} & c_{1}\chi_{1}^{*} & 0\end{array}\right) \, .
\end{equation}
Moreover the three characteristic velocities $c_n$  which appear in
(\ref{3wri}) are related to the six real constants $a_{n}$ and
$b_{n}$ by the relations  
\begin{equation}\label{c}
c_{n}=\frac{a_{n+1}-a_{n+2}}{b_{n+1}-b_{n+2}}\,,\quad 
n=1,2,3\,\,\,\textrm{mod}\left(3\right) .  
\end{equation}
The integrable character of the 3WRI equation (\ref{3wri}), together 
with
(\ref{gs}), follows from the compatibility ($\Psi_{xt}=\Psi_{tx})$
 of the two linear equations (\ref{laxpair}).
 
The system (\ref{3wri}), as it models the interaction of three 
resonating waves,
has attracted much attention in the study of nonlinear waves (see for instance 
\cite{Kaup}). It can 
be derived in
a quite general way via multiscale analysis of a large class of 
nonlinear wave
equations with weak dispersion and nonlinearity (see for instance \cite{Calogero}). As such, it 
represents a universal
model which is able to capture the lowest order corrections, due to 
nonlinear
effects, to linear propagation. It is here appropriate to point out
two special properties of the equation (\ref{3wri}) which make this 
model peculiar
with respect to most of the integrable nonlinear wave equations (f.i. 
Nonlinear
Schr\"{o}dinger equation, Korteweg--de Vries equation, Sine--Gordon 
equation and
others). The first one is that the system (\ref{3wri}) has no 
(linear) dispersion
(see (\ref{diffop})). The second property is that no 
self--interaction 
occurs, namely, the forcing of
each wave (i.e. the nonhomogeneous term in the propagation equation
(\ref{3wri})) is just the product of the (complex conjugated) 
amplitudes of the
other two waves, see the right hand side of (\ref{3wri}). If we 
assume that
the spacial profile of each one of the three waves is well localized 
in the far
past (i.e. as $t\rightarrow -\infty$), then these two properties 
imply that
each wave profile $\chi_n$ translates with no deformation with its 
characteristic
velocity $c_n$ up to the time it encounters an other wave. Then, for 
some time,
the interaction takes place where two waves (or the three of them) 
have
spacial overlapping, but eventually, in the far future (i.e. as 
$t\rightarrow
+\infty$), each wave is again free to move with no deformation with
its own characteristic velocity. Generally, the three wave profiles 
at $t=-\infty$
are different from those at $t=+\infty$ since the interaction at 
intermediate time
causes a change of the profile. In the $(x,t)$ plane the asymptotic 
behaviour is
therefore expressed by
\begin{equation}\label{asymp}
\lim_{t\rightarrow \pm \infty} \chi_n(y+ct,t)=0\,,\,\,\, c\neq 
c_n \,,\,\,\lim_{t\rightarrow \pm \infty} \chi_n(y+c_n t,t)= 
\chi_n^{(\pm)} (y)\,,
\end{equation}
where $y$ is the spacial coordinate in the moving reference frame, 
and the
functions $\chi_n^{(\pm)}(x-c_nt)$ are the asymptotic field profiles. 
Of course, the
relevant physical problem is computing the three wave packets in the 
future,  
$\chi_n^{(+)}(y)$, once the packets $\chi_n^{(-)}(y)$ in the past are
given. This large time asymptotic behaviour of the three waves may be 
referred to
as ``\emph{asymptotic freedom}". In terms of the spectral data 
associated to the three
functions $\chi_n(x,t)$ via the spectral problem 
$\Psi_{x}=(-ikB+U(x,t))\Psi$, 
see (\ref{laxpaira}), the spectral components of the asymptotic states
$\chi_n^{(\pm)}(y)$ may well  be on the continuum spectrum 
(wave--packets) other
than  on the discrete spectrum (solitons). These phenomena are quite 
in contrast
with the well--known ``\emph{separation}" process occurring for 
dispersive nonlinear
waves where only the discrete spectrum components (solitons) show up 
as
$t\rightarrow \pm
\infty$,  since the continuum spectrum components (wave packets) 
separate from solitons, disperse away and
vanish (see for instance \cite{CDeAC}). The asymptotic freedom behaviour of the three interacting 
waves may be given
a spectral formulation \cite{Kaup}, which is however not discussed 
here.

In Section 2 we collect formulae which provide us with the tool of the
Darboux--Dressing Transformation (DDT), while in Section 3 we obtain, 
via the DDT,
explicit solutions of the 3WRI equations (\ref{3wri}). In order to 
show the
asymptotic freedom for wave--packets, we deal with explicit solutions 
which
are not pure soliton solutions, this being the main result of this 
paper. An outlook with few remarks are the content of Section 4.
        
 \section{The Darboux--Dressing Transformation}
Let us now turn our attention to the method of construction of 
explicit 
solutions of the 3WRI equation (\ref{3wri}). This construction is 
based on the
Darboux transformation of ODEs and since this transformation is 
well-known \cite{DDT} we
limit our presentation below to a collection of formulae and 
propositions. To this
purpose it is convenient to observe first that, without any loss of 
generality, we
can give the constants $b_n$ and $a_n,\,\,n=1,2,3$ which appear in 
the Lax pair of
equations (\ref{laxpaira}) and, respectively, (\ref{laxpairb}), see 
also
(\ref{AB}), the following expressions
\begin{equation}\label{ab}
b_n=\frac{1}{3}(g_{n+1}-g_{n+2})\,,\,\,\,a_n=\frac{1}{3}(c_{n+1}g_{n+1}-
c_{n+2}g_{n+2})\,,\,\,\,n=1,2,3\,
,\,\,\textrm{mod}\left( 3\right)\, .
\end{equation}
The validity of this statement  follows from the expressions
(\ref{c}) of the characteristic velocities and (\ref{gs}) of the 
coupling
constants,  from the relation $b_1\,g_1+b_2\,g_2+b_3\,g_3=0$, which 
is implied by
them, and from  rescaling  the spectral parameter $k$. As a 
result, the spectral parameter $k$ has now the dimension of 
wave--number
$/$velocity. It should be also
noticed, for future reference, that the relations (\ref{ab}) can be 
inverted to
yield the expressions
\begin{equation}\label{bg}
g_n=-(b_{n+1}-b_{n+2}) \,,\,\,\,n=1,2,3\,
,\,\,\,\textrm{mod}\left( 3\right)\, ,
\end{equation}
which can be used as alternative to (\ref{gs}) with the implication 
that 
$b_n=-c_n+(c_1+c_2+c_3)/3$.
We  then note that both matrices
$U$ and
$V$ in the Lax pair (\ref{laxpair}) satisfy the reduction condition 
\begin{equation}\label{UVantih}
  U=-U^{\dagger}\, ,\quad V=-V^{\dagger}\, ,
\end{equation}
where the dagger denotes hermitian conjugation. 
This property (\ref{UVantih}) induces on the solution $\Psi$ 
of the two equations 
(\ref{laxpair}) the corresponding condition
\begin{equation}\label{psired}
\Psi^{\dagger}(x,t,k^\ast)\,\Psi(x,t,k)=C(k,k^\ast) \,,
\end{equation}
where the matrix $C(k,k^\ast)$ is constant, namely $x$-- and
$t$--independent; its value depends of course only on the 
arbitrary value $\Psi(x_0,t_0,k)$ that the solution
$\Psi$ takes at a  given point $(x_0,t_0)$ of the $(x,t)$ plane.

Consider now a second pair of off-diagonal matrices $U^{(0)}(x,t)$
and $V^{(0)}(x,t)$ which satisfy the same skew-Hermitian condition
$(U^{(0)})^{\dagger}=-U^{(0)}$, $(V^{(0)})^{\dagger}=-V^{(0)}$  as $U$
and $V$ (see (\ref{UVantih})); their entries, with obvious notation 
(see (\ref{UV})), identify the three complex functions  
 $\chi_{n}^{(0)}(x,t)$, $n=1,2,3$.
Let $\Psi^{(0)}(x,t,k)$ be a corresponding
nonsingular (i.e. with nonvanishing determinant) matrix solution of 
the
Lax pair (\ref{laxpair}), i.e.
\begin{equation}\label{0laxpair}
\Psi^{(0)}_x=(-ikB+U^{(0)})\Psi^{(0)}\, ,\quad
\Psi^{(0)}_t=(ikA+V^{(0)})\Psi^{(0)}
\end{equation} 
which have the same form (\ref{laxpair}) with $U$ and $V$ replaced,
respectively, by
$U^{(0)}$ and $V^{(0)}$. 
Assume that the initial condition $\Psi^{(0)}(x_0,t_0,k)$
is so chosen that the constant matrix $C^{(0)}(k,k^\ast)$ 
(see (\ref{psired})),
\begin{equation}\label{0psired}
C^{(0)}(k,k^\ast)={\Psi^{(0)}}^{\dagger}(x,t,k^\ast)\Psi^{(0)}(x,t,k)
\end{equation}
takes the same value of $C(k,k^\ast)$, i.e.
$C^{(0)}(k,k^\ast)=C(k,k^\ast)$.  
 If both compatibility conditions,
$\Psi^{(0)}_{xt}=\Psi^{(0)}_{tx}$ and $\Psi_{xt}=\Psi_{tx}$, are
satisfied, then $\chi_{n}^{(0)}$ and $\chi_{n}$, $n=1,2,3$, are two
different solutions of the same 3WRI equation (\ref{3wri}), and
 the matrix
\begin{equation}\label{Dmatrix}
D(x,t,k)=\Psi(x,t,k)(\Psi^{(0)}(x,t,k))^{-1} 
\end{equation}    
satisfies the pair of differential equations
\begin{subequations}
\label{Ddiff}
\begin{equation}\label{Ddiffa}
D_x=ik[D\,,B]+UD-DU^{(0)}\, , 
\end{equation} 
\begin{equation}\label{Ddiffb}
D_t=-ik[D\,,A]+VD-DV^{(0)}\, .
\end{equation} 
\end{subequations}
Moreover, as a consequence of the reduction conditions (\ref{psired}) 
and
(\ref{0psired}) with $C=C^{(0)}$, the matrix $D(x,t,k)$ satisfies 
also the
algebraic equation
\begin{equation}\label{Dalgeb}
D^\dagger(x,t,k^\ast)\,D(x,t,k)=I.
\end{equation} 
The proof of these statements is straight.

The definition (\ref{Dmatrix}) can be viewed as a transformation of
$\Psi^{(0)}$ into $\Psi$,
\begin{equation}\label{DDT}
\Psi(x,t,k)=D(x,t,k)\Psi^{(0)}(x,t,k) \, ,
\end{equation}
which consequently yields a transformation of $\chi_{n}^{(0)}$ (the
\emph{bare} solution) into $\chi_{n}$ (the \emph{dressed} solution). 
Clearly the dressing approach requires in the first place that
$\chi_{n}^{(0)}(x,t)$ and $\Psi^{(0)}(x,t,k)$ be explicitly known. 
The next
step is the construction of the transformation matrix $D(x,t,k)$ via
the integration of the ODEs (\ref{Ddiff}). This task however is not
straight since the coefficients of these differential
equations depend on the unknown matrices $U$ and $V$, see 
(\ref{Ddiffa})
and (\ref{Ddiffb}). The way to overcome this difficulty goes through
the \emph{a priori} assignement of the dependence of the
transformation matrix $D(x,t,k)$ on the spectral variable $k$. 

The simplest instance of this strategy is illustrated by the following
proposition (whose proof is a mere exercise): if $D$ is $k$--independent,
$D_k=0$, then $D$ is diagonal, $[A ,D]=0$, $[B ,D]=0$, and is $x$--and
$t$--independent, i.e. $D_x=D_t=0$.  The associated transformation of
$U^{(0)}$ is then a gauge transformation, namely $D=G$,
$U=GU^{(0)}G^{-1}$.

In the following we will consider the larger
set of $k$--dependent matrices $D(k)$ which (i) have a
\emph{rational} dependence on the complex variable $k$ and (ii) have
\emph{nonvanishing} limit as $k\rightarrow \infty$ . As we consider here
 only rational dependence on $k$ which can be factorized as product of
simple-pole terms, we need to deal only with matrices $D(x,t,k)$ which
take the following one-pole expression
\begin{equation}\label{DDTmat}
D(x,t,k)=\mathbf{1}+\frac{R(x,t)}{k-\alpha}\,,
\end{equation} 
the matrix $R(x,t)$ being the residue at the pole $k=\alpha$. It is 
plain
that the transformation matrices $D(k)$ whose $k\rightarrow \infty$ 
limit
is a nonsingular, but not necessarily unit, matrix may be obtained by 
multiplying
 the expression (\ref{DDTmat}) by a gauge transformation $G$. The
transformation  characterized by the matrix (\ref{DDTmat}) has 
received
considerable attention in the literature \cite{DDT}; 
we refer to it as  to Darboux--Dressing
Transformation (DDT). Its existence in our setting is proved below by
construction.

In general, the way to obtain an explicit expression of the residue 
matrix $R(x,t)$
depends on whether the pole $\alpha$ is on the real axis,
$\alpha=\alpha^\ast$, or not. However, in the case under 
investigation, it is easily seen that the condition 
$\alpha=\alpha^\ast$ implies
$R(x,t)=0$.  Thus we need to consider  only the case in
which $\alpha$ is not real, $\alpha\neq\alpha^\ast$.  The starting
point is the requirement that the matrix $D(x,t,k)$
satisfies the algebraic condition (\ref{Dalgeb}) and
the differential equations (\ref{Ddiff}). The algebraic condition 
(\ref{Dalgeb}) 
entails the equation
\begin{equation}\label{algeq}
R+\frac{R^\dagger R}{\alpha-\alpha^\ast}=0\, ,
\end{equation}  
whose solution is 
\begin{equation}\label{residue}
R(x,t)=(\alpha-\alpha^\ast)P(x,t)\, ,
\end{equation} 
where the matrix $P(x,t)$ is an Hermitian projector
\begin{equation}\label{proj}
P^2=P\,,\quad P^{\dagger}=P\,\, .
\end{equation}
As for the two differential equations (\ref{Ddiff}), replacing 
$D(x,t,k)$
with its expression
\begin{equation}\label{elementaryDDT}
D(x,t,k)=\mathbf{1}+\left(\frac{\alpha-\alpha^\ast}{k-\alpha}\right) 
P(x,t)\,,
\end{equation}
(see (\ref{DDTmat}) and (\ref{residue})) yields two algebraic and two 
differential equations.
The algebraic relations read
\begin{gather}\label{Backlund}
U=U^{(0)}+i(\alpha-\alpha^\ast)[B\,,P]\, ,\\
\nonumber\\
V=V^{(0)}-i(\alpha-\alpha^\ast)[A\,,P]\, ,
\end{gather} 
and give the \emph{dressed} quantities $\chi_{n}(x,t)$, in terms of 
the 
\emph{bare} ones
$\chi_{n}^{(0)}(x,t)$ $n=1,2,3$,  and the projector $P$ (see below). 
 The two differential
equations are
\begin{subequations} 
\label{diffeq}
\begin{equation}\label{diffeqa}
P_x=(1-P)(-i\alpha^{*}B+U^{(0)})P-P(-i\alpha B+U^{(0)})(1-P)\,,
\end{equation}
\begin{equation}\label{diffeqb}
P_t=(1-P)(i\alpha^{*}A+V^{(0)})P-P(i\alpha A+V^{(0)})(1-P)\, .
\end{equation}
\end{subequations}
Let $v$ be an eigenvector of P with unit eigenvalue, and 
differentiate 
with respect to $x$ the eigenvalue equation 
\begin{equation}\label{eigenvector}
Pv=v\, ,\quad 
v=\left(
\begin{array}{c}
    v_1\\ v_2\\ v_3 \end{array} \right)\, .
\end{equation}
By replacing then $P_x$ with the right-hand side of (\ref{diffeqa}), 
one
obtains the equation
\begin{equation}\label{vxdiff}
(\mathbf{1}-P)[v_x-(-i\alpha^{*}B+U^{(0)})v]=0\, ,
\end{equation}
which implies that the vector $v_x-(-i\alpha^{*}B+U^{(0)})v$ belongs 
to 
the subspace on which $P$ projects. At this point we note that, since 
the dimension of
$P$ is $3$, we may well assume, without any loss of generality, 
that this subspace
be one--dimensional. Indeed, it is easy to show that if $P$ projects 
on a
two--dimensional subspace, then $\mathbf{1}-P$ projects on a 
one--dimensional
subspace and the only change in the final result is the value of the 
parameter $\alpha$, which goes into
$\alpha^{\ast}$.  Therefore, 
$P$ projects on the one--dimensional
subspace of the vector $v$:
\begin{equation}\label{diadic}
P(x,t)=\frac{v(x,t)v^\dagger(x,t)}{v^{\dagger}(x,t)v(x,t)}\,\,,
\end{equation}
where in this notation we treat vectors as one-column rectangular 
matrices.
This implies, see (\ref{vxdiff}), that the vector
$v_x-(-i\alpha^{*}B+U^{(0)})v$ is proportional to $v$, i.e.
\begin{equation}\label{vxdiff1}
v_x-(-i\alpha^{*}B+U^{(0)})v=h v\, .
\end{equation}
 On the other hand,
since the vector $v$ is identified only modulo a scalar factor, which 
may 
well be a  function of $x$ and $t$, we
may choose this factor in such a way that $v$ satisfies this 
differential
equation (\ref{vxdiff1}) with $h=0$.

The differential equation (\ref{diffeqb}) with respect to the variable
$t$ can be handled in a similar way. 
One obtains therefore that the vector $v(x,t)$
satisfies the differential equations
\begin{subequations}
\label{zeqs}
\begin{equation}\label{zeqsa}
v_x=(-i\alpha^{*}B+U^{(0)})v\, , 
\end{equation}
\begin{equation}\label{zeqsb}
v_t=(i\alpha^{*}A+V^{(0)})v\,.
\end{equation}
\end{subequations}
Once the two equations (\ref{zeqs}) are solved, the DDT
transformation matrix $D(x,t,k)$ is finally given explicitly by
(\ref{elementaryDDT}) with (\ref{diadic}).

At this point, we conclude that the method of construction of a novel
solution $\chi_{n}(x,t)$ of the 3WRI equation (\ref{3wri}), starting 
from
the knowledge of a given  solution $\chi^{(0)}_{n}(x,t)$ is 
explicitly given
by (\ref{Backlund}) with  (\ref{diadic}) where the
vector $v(x,t)$ is given by
\begin{equation} \label{vsolution}
v(x,t)=\Psi^{(0)}(x,t,\alpha^\ast)\,v_0\,.
\end{equation}
Here $\Psi^{(0)}(x,t,\alpha^\ast)$ is assumed to be a known solution
$\Psi^{(0)}(x,t,k)$  of the Lax pair 
(\ref{0laxpair}), for $k=\alpha^\ast$, while $v_0$ is an arbitrary 
constant vector,
\begin{equation}\label{z0}
v_0=\left(\begin{array}{c}\gamma_1\\ \gamma_2\\ \gamma_3 \end{array} 
\right)\,\,,
\end{equation}
where $\gamma_1,\gamma_2,\gamma_3$ are three (arbitrary) complex 
parameters. With
these specifications (see also (\ref{bg})), the DDT transformation
(\ref{Backlund}) takes the more explicit form
\begin{equation}\label{explDDT}
\chi_n=\chi_n^{(0)}-i(\alpha-\alpha^\ast)g_n\frac{v_{n+1}
v_{n+2}^{\ast}}{|v_1|^2+|v_2|^2+|v_3|^2}\, ,\quad 
n=1,2,3\,,\,\,\,\textrm{mod}\left(3\right)\,.
\end{equation}

\section{Solutions}\label{solutions}
In this section we apply the DDT to known solutions 
$\chi_n^{(0)}\,(x,t)$ of the
3WRI equation (\ref{3wri}) to obtain other solutions of the same 
equation. Let us 
consider first the DDT within the class of solutions $\chi_n(x,t)$ of 
the 3WRI system
(\ref{3wri}) which are well localized at all times. This class is 
left invariant by 
the DDT if the functions $\chi_n(x,t)$ vanish sufficiently fast for 
large $|x|$ at 
fixed $t$, $\lim_{x\rightarrow \pm \infty}\chi_n(x,t)=0$ (for 
instance, if they are 
in $L_1(R)$ as functions of the 
$x$ variable). The simplest choice of the bare solution
$\chi_n^{(0)}\,(x,t)$ in this class is, of course, the vanishing one, 
i.e.
$\chi_n^{(0)}\,(x,t)=0$. The corresponding solution which is obtained by 
applying to it 
the DDT described in the previous section is the well-known 
one-soliton solution 
(see below). The more general class of solutions we compute here 
originates 
from the simple observation that  even asking that  only two of the 
three fields 
$\chi_1 ,  \chi_2$ and $\chi_3$ are 
vanishing yields a solution, which depends moreover on an arbitrary 
complex 
function of one real variable. Since the non vanishing field can be 
any one of the
three, we introduce three different bare solutions as distinguished 
by the index $j$:
\begin{equation}\label{bare}
\chi_{n}^{(0,j)}(x,t)=\delta_{j n} f(x_n) \,, \,\, j=1,2,3\,, 
\end{equation}
where $f(y)$ is an arbitrary complex function of the real variable 
$y$ and the three 
variables $x_n$, $\,n=1,2,3$, are the  \emph{characteristic 
coordinates} 
\begin{equation}\label{xn}
x_{n}\equiv x-c_{n}\,t,\quad n=1,2,3\,.  
\end{equation}
\emph{Remark}. The distinction of the three solutions (\ref{bare}) by 
the index $j$ 
makes sense if  the labeling  of the three fields $\chi_n$ by the 
index $n$ is fixed. 
From now on, without any loss of generality, we set such labeling by 
ordering 
the three characteristic velocities in the following way:
\begin{equation}\label{order}
c_1~<~c_2~<c_3\,\,\, .
\end{equation} 
Indeed we remind the reader that all other relevant parameters, 
namely the 
coupling constants $g_n$, see (\ref{gs}), and the constants $a_n$ and 
$b_n$, 
see (\ref{ab}), which appear in the Lax equations, see 
(\ref{laxpaira}) and 
(\ref{laxpairb}) with (\ref{AB}), are functions of the $c_n$'s. 
 
The first step in applying the DDT to the seed solution (\ref{bare}) is of 
course solving the Lax pair of equations 
(\ref{0laxpair}) for 
each one of the three cases (with obvious notation, and see(\ref{UV}))
\begin{equation}\label{bareUV}
    \begin{array}{ll}
	U^{(0,1)}=\left( \begin{array}{ccc}
0 & 0 & 0\\
0 & 0 & f(x_1)\\
0 & -f^{*}(x_1) & 0\end{array}\right)\, ,&
\,\,U^{(0,2)}=\left( \begin{array}{ccc}
0 &0 & -f^{*}(x_2)\\
0 & 0 & 0\\
f(x_2) & 0 & 0\end{array}\right) \, ,\\
\\
U^{(0,3)}=\left( \begin{array}{ccc}
0 &f(x_3)& 0\\
-f^{*}(x_3) & 0 & 0\\
0 & 0 & 0\end{array}\right)\, ,&  \,\,V^{(0,j)}=-c_j U^{(0,j)}\, ,
\end{array}
 \end{equation}
which correspond to the three bare solutions (\ref{bare}). Because of 
the 
particular structure of the  matrices $U^{(0,j)}$, see 
(\ref{bareUV}), and 
of their dependence on $x$ and $t$ through the characteristic 
coordinates, 
see (\ref{xn}), the matrix solution $\Psi^{(0,j)}(x,t,k)$ takes the 
form
\begin{equation}\label{barepsi}
\Psi^{(0,j)}(x,t,k)=\Phi^{(j)}(x_j,k) \exp\left[i\frac{k}{2} (b_{j} x - 
a_{j} t)
(\mathbf{1}-3P_j)\right]\,\,,
\end{equation}   
where the three matrices $P_j\,~~j=1\,,2\,,3,~$ are the three 
diagonal projectors
\begin{equation}\label{diagprojectors}
P_1=\left(\begin{array}{ccc}
1 & 0 & 0\\0 &  0 & 0\\0 &  0 & 
0\end{array}\right)\,,\quad P_2=\left(\begin{array}{ccc}
0 & 0 & 0\\0 &  1 & 0\\0 &  0 & 
0\end{array}\right)\,,\quad P_3=\left(\begin{array}{ccc}
0 & 0 & 0\\0 &  0 & 0\\0 &  0 & 1\end{array}\right)\,\,.
\end{equation}
Solving the equations of the Lax pair now reduces to solving only the 
$3\times3$ matrix ODE
\begin{equation}\label{phieq}
\Phi^{(j)}_{y} (y,k)=\left[\frac{i}{2} g_j k(P_{j+1}-P_{j+2}) + 
 U^{(0,j)}(y)\right]\Phi^{(j)} (y,k)\,\,,
\end{equation}
which is however easily seen to be equivalent to the standard  
Zakharov-Shabat  
system of two equations
\begin{equation}\label{ZS}
\phi_{1y}=\frac{i}{2}\lambda \phi_1 + f(y)\phi_2\,\,,\,\,\,\phi_{2y}=
-\frac{i}{2}\lambda \phi_2 - f^{*}(y)\phi_1
\end{equation}
for the two functions $\phi_1(y,\lambda)$ and $\phi_2(y,\lambda)$. 
Indeed, once a solution of the system (\ref{ZS}) is known, the 
$3\times3$ matrices $\Phi^{(j)}(y,k)$, for $j=1\,,\,2\,,\,3\,,$ 
which solve (\ref{phieq}) are also known and read
\begin{subequations}
\begin{equation}\label{phimatrix1}
\Phi^{(1)} (y,k)=\left(\begin{array}{ccc}1 & 0 & 0\\
0 & \phi_1(y,g_1 k) & -\phi_2^* (y,g_1 k^*)\\
0 &  \phi_2(y,g_1 k) & \phi_1^* (y,g_1 k^*)\end{array}\right) \,\,,
\end{equation}
\begin{equation}\label{phimatrix2}
\Phi^{(2)} (y,k)=\left(\begin{array}{ccc}\phi_2(y,g_2 k) & 0 & 
-\phi_1^* (y,g_2 k^*)
\\0 &  1 & 0\\\phi_1(y,g_2 k)&  0 & \phi_2^* (y,g_2 
k^*)\end{array}\right) \,\,,
\end{equation}
\begin{equation}\label{phimatrix3}
\Phi^{(3)} (y,k)=\left(\begin{array}{ccc}\phi_1(y,g_3 k) & -\phi_2^* 
(y,g_3 k^*) & 0
\\\phi_2(y,g_3 k) &  \phi_1^* (y,g_3 k^*) & 0\\0 &  0 & 
1\end{array}\right)\,\,.
\end{equation}
\end{subequations}
For future reference, we display also the asymptotic behaviour of the 
solution 
of the Zakharov-Shabat  system (\ref{ZS})
\begin{equation}\label{z12}
\phi_1(y,\lambda)\underset{y\to \pm 
\infty}{\longrightarrow}z_{1}^{(\pm)}
(\lambda) \exp\left(\frac{i}{2}\lambda y\right)\,\,, \,\,\,\phi_2(y,\lambda)
\underset{y\to \pm \infty}{\longrightarrow}z_{2}^{(\pm)}(\lambda)
 \exp\left(-\frac{i}{2}\lambda y\right)\,\,, 
\end{equation}
which introduces the four functions $z_{1}^{(\pm)}(\lambda)$ and 
$z_{2}^{(\pm)}(\lambda)$, two of which can of course be arbitrarily 
given while the other two can be obtained by integrating the ODEs (\ref{ZS}). 

Let us now consider the dressed solution by applying the DDT given in  
the previous section to the bare solution (\ref{bare}). Its 
expression 
(see (\ref{explDDT})) is
\begin{equation}\label{dressed}
\chi_n^{(j)}(x,t)=\delta_{j n} f(x_n)+2\eta 
g_n\frac{v^{(j)}_{n+1}(x,t)
v^{(j)*}_{n+2}(x,t)}{|v^{(j)}_1(x,t)|^2+|v^{(j)}_2(x,t)|^2+|v^{(j)}_3(x,t)|^2}\,\,\,,
\end{equation}
where  the parameter $\eta$ is the imaginary part of the complex pole 
$\alpha$ which has been introduced in the DDT formula 
(\ref{elementaryDDT}), namely
\begin{equation}\label{pole}
\alpha=\rho+i\eta \,\,.
\end{equation}
According to the definition (\ref{vsolution}), the three 3-vectors 
$v^{(j)}(x,t)$, 
 with components $v_n^{(j)}(x,t)$, are provided by the formula 
(see(\ref{barepsi}))
\begin{equation}\label{vj}
v^{(j)}(x,t)=\Phi^{(j)}(x_j,\alpha^*) \exp\left[i\frac{\alpha^*}{2} (b_{j} 
x - a_{j} t)(\mathbf{1}
-3P_j)\right] v_0\,\,,
\end{equation}
where the arbitrary complex constant vector $v_0$ introduces in the 
solution 
$\chi_n^{(j)}(x,t)$ the three arbitrary parameters 
$\gamma_1\,,\,\gamma_2 \,$and\,
 $\gamma_3$, see (\ref{z0}). This new solution (\ref{dressed}) of the 
3WRI 
equation depends on the function $f(y)$ which characterizes the bare 
solution
 (\ref{bare}) and its behavior may be roughly described as  the 
interaction 
of the bare $j$-th wave  $f(x-c_j t)$ with  soliton--type bumps in the 
other
 two waves whose speed may be larger or smaller than $c_j$, a 
behavior which 
may be quite complicate at finite intermediate times. Here we do not 
dwell 
in detailing such behavior at finite times, rather we compute
 the asymptotic states as $t\rightarrow \pm \infty$  
whose profile 
and properties really matter in an applicative context. According to 
the general 
definition (\ref{asymp}), we now compute, for each value of $j$, the 
six functions
 $\chi^{(j)(\pm)}_n(y)$ by performing the limits
\begin{equation}
\lim_{t\rightarrow \pm \infty} \chi^{(j)}_n(y+c_nt,t)= 
\chi_n^{(j)(\pm)} (y)
\,\,\,.
\end{equation}
This computation is elementary but lengthy, and it is omitted. The 
relevant 
results take the following expressions:
\begin{subequations}
\begin{equation}\label{limitj}
\left\{ \begin{array}{l}\lim_{t\rightarrow s_j \infty} 
\chi^{(j)}_j(y+c_jt,t)=f(y)+2\eta g_j \frac{w^{(j)}_{j+1} (y) w^{(j)*}_{j+2} (y)}
{|w^{(j)}_{j+1} (y)|^2 +|w^{(j)}_{j+2} (y)|^2}\,,\\
\lim_{t\rightarrow -s_j \infty} \chi^{(j)}_j(y+c_jt,t)=f(y)\,\,,
\end{array}
\right. 
\end{equation} 
\begin{equation}\label{limitj+1}
 \left \{\begin{array}{l} \lim_{t\rightarrow s_{j+1} \infty} 
\chi^{(j)}_{j+1}(y+c_{j+1}t,t)=\\
\hspace{3cm}=2\eta g_{j+1} \exp(i\rho g_{j+1} y)
 \frac{\gamma^*_j u^{(j)(-s_j)}_{j+2}}{|\gamma_j|^2 \exp(-\eta 
g_{j+1} y)
+| u^{(j)(-s_j)}_{j+2}|^2  \exp(\eta g_{j+1} y)} \,,\\ 
\lim_{t\rightarrow 
- s_{j+1} \infty} \chi^{(j)}_{j+1}(y+c_{j+1}t,t)=0\,\,,
\end{array}
\right.  
\end{equation}
\begin{equation}\label{limitj+2}
 \left \{\begin{array}{l} \lim_{t\rightarrow s_{j+2} \infty} 
\chi^{(j)}_{j+2}(y+c_{j+2}t,t)=\\
\hspace{3cm}=2\eta g_{j+2} \exp(i\rho g_{j+2} y)
 \frac{\gamma_j u^{(j)(s_j)*}_{j+1}}{|\gamma_j|^2 \exp(\eta g_{j+2} y)
+| u^{(j)(s_j)}_{j+1}|^2  \exp(-\eta g_{j+2} y)} \,\,, \\ 
\lim_{t\rightarrow -s_{j+2} \infty} 
\chi^{(j)}_{j+2}(y+c_{j+2}t,t)=0\,\,.
\end{array}
\right.  
\end{equation}
\end{subequations}
The symbols we have introduced here are:
the three signs
\begin{equation}\label{signs}
s_j= \mathrm{sign}(\eta g_j)\,,\,\,\, s_j^2=1\,,\,\,\,j=1\,,2\,,3\,,
\end{equation}
the real parameter $\rho$ which is the real part of the complex pole 
$\alpha$, see (\ref{pole}),
the three $y-$dependent vectors (see (\ref{phimatrix1}, 
\ref{phimatrix2},
 \ref{phimatrix3}) and (\ref{z0}))
\begin{equation}\label{wvector}
w^{(j)}(y)=\Phi^{(j)}(y,\alpha^*) v_0\,\,,
\end{equation}
and the three constant vectors
\begin{subequations}
\begin{equation}\label{u1vector}
u^{(1)(\pm)}=\left(\begin{array}{c}
\gamma_1 \\\gamma_2 z_{1}^{(\pm)}(g_1 \alpha^*)-\gamma_3 
z_{2}^{(\pm)*}(g_1 \alpha)
\\ \gamma_2 z_{2}^{(\pm)}(g_1 \alpha^*)+\gamma_3 z_{1}^{(\pm)*}(g_1 
\alpha)
\end{array}\right)\,\,,
\end{equation}
\begin{equation}\label{u2vector}
u^{(2)(\pm)}=\left(\begin{array}{c}
\gamma_1 z_{2}^{(\pm)}(g_2 \alpha^*)-\gamma_3 z_{1}^{(\pm)*}(g_2 
\alpha)
\\\gamma_2\\\gamma_1 z_{1}^{(\pm)}(g_2 \alpha^*)
+\gamma_3 z_{2}^{(\pm)*}(g_2 \alpha)\end{array}\right)\,\,,
\end{equation}
\begin{equation}\label{u3vector}
u^{(3)(\pm)}=\left(\begin{array}{c}
\gamma_1 z_{1}^{(\pm)}(g_3 \alpha^*)-\gamma_2 z_{2}^{(\pm)*}(g_3 
\alpha)
\\\gamma_1 z_{2}^{(\pm)}(g_3\alpha^*)+\gamma_2 z_{1}^{(\pm)*}(g_3 
\alpha)
\\\gamma_3\end{array}\right)\,\,,
\end{equation}
\end{subequations}
where the four parameters $z_{1}^{(\pm)} (g_j \alpha)$,
 $z_{2}^{(\pm)} (g_j \alpha)$ and $z_{1}^{(\pm)} (g_j \alpha^*)$, 
$z_{2}^{(\pm)} (g_j \alpha^*)$, for $j=1,2,3$, are the values that 
the 
two functions $z_{1}^{(\pm)}(\lambda)$ and $z_{2}^{(\pm)}(\lambda)$, 
introduced via the asymptotic behavior (\ref{z12}), take at  
 $\lambda=g_j \alpha$ and $\lambda=g_j \alpha^*$, respectively. 

At this point we are in the position to discuss the asymptotic in- 
and out-states corresponding to our solution. Since we aim to attract 
the attention of the reader on potential applications, we chose to 
have 
the bare wave $f(x-c_j t)$ in the initial state (say as $t\rightarrow 
-\infty$).
 Therefore, as indicated by the limit formula (\ref{limitj}), we have 
to set $s_j=1$ for each $j$, and this implies, according to the 
definition (\ref{signs}) and to the signs of the coupling constants, 
see (\ref{gs}) and (\ref{order}),
\begin{equation}\label{gsign}
\mathrm{sign}(g_1)=-1\,\,,\,\,\mathrm{sign}(g_2)=1\,\,,\,
\,\mathrm{sign}(g_3)=-1\,\,,
\end{equation}
that we pick the pole $\alpha$ of the DDT in the lower half of 
the complex plane, $\eta<0$, for $j=1$ and $j=3$ and in the upper 
half of the complex plane, $\eta>0$, for $j=2$. By taking into 
account 
that these choices of the sign of the parameter $\eta$ entail that 
\begin{equation}\label{jsigns}
\left\{ \begin{array}{l}s_2=-1\,,\,\,\,s_3=1\,,\,\,\,\mathrm{for}\,\,j=1\,
\,,\\
s_1=-1\,,\,\,\,s_3=-1\,,\,\,\,\mathrm{for}\,\,j=2\,\,,\\
s_1=1\,,\,\,\,s_2=-1\,,\,\,\,\mathrm{for}\,\,j=3\,\,,\,
\end{array}
\right. 
\end{equation}
the final expressions of the asymptotic profiles of the three waves 
are 
readily read out of the formulae (\ref{limitj}), (\ref{limitj+1}) and 
(\ref{limitj+2}), and are
\begin{subequations}
\begin{equation}\label{1asymp}
\chi^{(1)(-)}(x, t)=\left(\begin{array}{c}f(x-c_1 t) 
\\S_2^{(1)}(x-c_2 t)
\\0\end{array}\right)\,,\,\,
\chi^{(1)(+)}(x, t)=\left(\begin{array}{c}F^{(1)}(x-c_1 t) \\0
\\S_3^{(1)}(x-c_3 t)\end{array}\right)\,,
\end{equation}
\begin{equation}\label{2asymp}
\chi^{(2)(-)}(x, t)=\left(\begin{array}{c}
S_1^{(2)}(x-c_1 t) \\f(x-c_2 t)\\S_3^{(2)}(x-c_3 
t)\end{array}\right)\,
,\,\,\chi^{(2)(+)}(x, t)=\left(\begin{array}{c}
0\\F^{(2)}(x-c_2 t)\\0\end{array}\right)\,,
\end{equation}
\begin{equation}\label{3asymp}
\chi^{(3)(-)}(x, t)=\left(\begin{array}{c}
0\\S_2^{(3)}(x-c_2 t)\\f(x-c_3 
t)\end{array}\right)\,,\,\,\chi^{(3)(+)}(x, t)
=\left(\begin{array}{c}
S_1^{(3)}(x-c_1 t)\\0\\F^{(3)}(x-c_3 t)\end{array}\right)\,.
\end{equation}
\end{subequations}
In these formulae we have expressed the asymptotic profiles by means 
of 
the following expressions, which again are just implied by 
(\ref{limitj}), 
(\ref{limitj+1}) and (\ref{limitj+2}):
\begin{equation}\label{F}
F^{(j)}(y)=f(y)+2\eta g_j \frac{w^{(j)}_{j+1} (y) w^{(j)*}_{j+2} (y)}
{|w^{(j)}_{j+1} (y)|^2 +|w^{(j)}_{j+2} (y)|^2}\,,\,
\end{equation}
\begin{subequations}
\begin{equation}\label{1soliton}
\left\{ \begin{array}{l}S_2^{(1)}(y)=2\eta g_{2} \exp(i\rho g_{2} y) 
\frac{\gamma^*_1 u^{(1)(-)}_{3}}{|\gamma_1|^2 \exp(-\eta g_{2} y)
+| u^{(1)(-)}_{3}|^2  \exp(\eta g_{2} y)} \,\,,\\
S_3^{(1)}(y)=2\eta g_{3} \exp(i\rho g_{3} y) \frac{\gamma_1 
u^{(1)(+)*}_{2}}
{|\gamma_1|^2 \exp(\eta g_{3} y)+| u^{(1)(+)}_{2}|^2  \exp(-\eta 
g_{3} y)} \,
\,,
\end{array}
\right. 
\end{equation} 
\begin{equation}\label{2soliton}
 \left\{ \begin{array}{l}S_1^{(2)}(y)=2\eta g_{1} \exp(i\rho g_{1} y) 
\frac{\gamma_2 u^{(2)(+)*}_{3}}{|\gamma_2|^2 \exp(\eta g_{1} y)+
| u^{(2)(+)}_{3}|^2  \exp(-\eta g_{1} y)} \,\,,\\
S_3^{(2)}(y)=2\eta g_{3} \exp(i\rho g_{3} y) \frac{\gamma^*_2 
u^{(2)(-)}_{1}}
{|\gamma_2|^2 \exp(-\eta g_{3} y)+| u^{(2)(-)}_{1}|^2  \exp(\eta 
g_{3} y)} \,\,,
\end{array}
\right. 
\end{equation}
\begin{equation}\label{3soliton}
 \left\{ \begin{array}{l}S_1^{(3)}(y)=2\eta g_{1} \exp(i\rho g_{1} y) 
\frac{\gamma^*_3 u^{(3)(-)}_{2}}{|\gamma_3|^2 \exp(-\eta g_{1} y)
+| u^{(3)(-)}_{2}|^2  \exp(\eta g_{1} y)}\, \,,\\
S_2^{(3)}(y)=2\eta g_{2} \exp(i\rho g_{2} y) \frac{\gamma_3 
u^{(3)(+)*}_{1}}
{|\gamma_3|^2 \exp(\eta g_{2} y)+| u^{(3)(+)}_{1}|^2  \exp(-\eta 
g_{2} y)} \,\,.
\end{array}
\right. 
\end{equation}  
\end{subequations}
Their meaning  is quite transparent: the function $F^{(j)}(y)$, which 
always appears in an out-state, shows how the profile of the 
incoming wave--packet $f(y)$ changes because of the collision with 
solitonic waves, while the functions $S^{(j)}_n (y)$ show the 
standard 
sech--shape of the solitons (see below) appearing in the incoming 
and/or 
outgoing states.
The three different collision processes described by these asymptotic 
states,
 (\ref{1asymp}), (\ref{2asymp}), (\ref{3asymp}), are controlled by 
various 
parameters introduced via the DDT, namely the complex 
pole
 $\alpha=\rho+i\eta$  and the complex components $\gamma_j$ of the 
vector
 $v_0$ (see (\ref{z0})) on one hand, and the (largely arbitrary) 
complex 
function $f(y)$ characterizing the bare solution on the other hand. 
We note moreover that only in the process with $j=2$ there are three 
non vanishing waves in the in-state, a feature which is not found in 
the one-soliton solution (see below).  We also note that not only the 
deformation  $F^{(j)}(y)$ of the initial profile $f(y)$
 (see (\ref{wvector})) depends on the solution of the ZS
 equations $(\ref{ZS})$, but also the position of the in-coming 
and out-going solitons (see below) depends on these solutions through 
their asymptotic behavior, see (\ref{1soliton}), (\ref{2soliton}),
 (\ref{3soliton}) together with (\ref{u1vector}), (\ref{u2vector}), 
(\ref{u3vector}) and (\ref{z12}). 

At this point it is evident that numerical experiments, which 
do not rely on the integration of the 3WRI system of PDEs (\ref{3wri})
 but just on analytical explicit expressions, are possible if one 
chooses 
the function $f(y)$ as a ``solvable potential", namely  a function 
such that 
the general solution of the  ZS equations (\ref{ZS}) is explicitly 
known. 
There exist indeed many such choices. We just pick one of them here 
below 
to produce explicit formulae which one can use to experiment on 
three--wave
 interaction processes. Before doing this however we observe that
the asymptotic profiles $S^{(j)}_n(y)$, see (\ref{1asymp}), 
(\ref{2asymp}),
 (\ref{3asymp}) and  (\ref{1soliton}), (\ref{2soliton}), 
(\ref{3soliton}),
 coincide with the asymptotic states of the one-soliton solution of 
the 
3WRI equation. This solution obtains by applying the DDT to the 
trivial
 choice $f(y)=0$ and by using the formulae given above. In this 
particular
 case the index $j$ which we use to distinguish three different 
initial
 bare solutions (see (\ref{bare})) is, of course, useless and it is 
dropped.
 The resulting expression of the one-soliton solution then reads, in 
our notation,
{\small \begin{equation}\label{standardsoliton1}
    \begin{array}{l}
\chi_{n}(x,t)=2\eta g_{n}\,\exp(i\rho g_{n} x_{n})\times\\
\times\frac{\gamma_{n+1}\,\gamma_{n+2}^{\ast}}{|\gamma_{n+1}|^{2}\,
\exp(\eta g_{n}\,x_{n})+
|\gamma_{n+2}|^{2}\exp(-\eta g_{n}\,x_{n})+|\gamma_{n}|^{2}\,
\exp[\eta (g_{n+2}-g_{n+1})\,x_{n}] \exp(-2\eta 
g_{n+1}g_{n+2}\,t)} \,\, , 
\end{array}
\end{equation}}
where the signs $s_n\,,\,n=1\,,\,2\,,\,3\,,$ are defined by 
(\ref{signs})
 with the order convention (\ref{order}). The large time limits of 
this
 solution are readily found to be
\begin{subequations}
\begin{equation}\label{asympt1}
\chi_{n}(y+c_{n}\,t,t)\underset{t\to -s_{n}\infty}{\longrightarrow}0\, ,
\end{equation}
\begin{equation}\label{asympt2}
\chi_{n}(y+c_{n}\,t,t)\underset{t\to s_{n}\infty}{\longrightarrow}
2\eta g_{n}\,\exp(i\rho g_{n} y)\frac{\gamma_{n+1}\,\gamma
_{n+2}^{\ast}}{|\gamma_{n+1}|^{2}\,
\exp(\eta g_{n}\,y)+
|\gamma_{n+2}|^{2}\exp(-\eta g_{n}\,y)}\, ,
\end{equation}
\end{subequations}
and this last expression coincides, apart from obvious replacement of 
the
 parameters $\gamma$'s, with the general expressions (\ref{limitj})
 with $f(y)=0$ or (\ref{limitj+1}) or (\ref{limitj+2}). If we choose 
moreover $\eta >0$, it is easily seen that, because of (\ref{gsign}),
 the asymptotic states here correspond to the expressions given above 
for $j=2$ and $f(y)=0$. These states may be  given also the familiar 
sech form
\begin{subequations}
\begin{equation}\label{-soliton}
\left\{ \begin{array}{l}\chi_1^{(-)}(y)=\frac{\eta \,g_{1}\,
\exp \left[ i\,\left( \rho
\,g_{1}\,y+\theta _{2}-\theta _{3}\right) \right]}{\cosh \left[ \eta
\,g_{1}\left(y-\xi_{1}\right) \right]}\,,\\ 
\chi_2^{(-)}(y)=0\,,\\
\chi_3^{(-)}(y)=\frac{\eta \,g_{3}\,\exp \left[ i\,\left( \rho
\,g_{3}\,y+\theta _{1}-\theta _{2}\right) \right]}{\cosh \left[ \eta
\,g_{3}\left(y-\xi_{3}\right) \right]}\,,
\end{array}
\right. 
\end{equation} 
\begin{equation}\label{+soliton}
 \left \{\begin{array}{l} \chi_1^{(+)}(y)=0\,,\\ 
\chi_2^{(+)}(y)=\frac{\eta
\,g_{2}\,\exp
\left[ i\,\left( \rho
\,g_{2}\,y+\theta _{3}-\theta _{1}\right) \right]}{\cosh \left[ \eta
\,g_{2}\left(y-\xi_{2}\right) \right]}\,,\\ 
\chi_3^{(+)}(y)=0,
\end{array}
\right. 
\end{equation}
\end{subequations}
by introducing the phases $\theta_n$ and the positions $\xi_n$
 via the relation
\begin{equation}\label{theta}
\gamma_{n}=\exp(i\theta_{n})\,\exp\left[\frac{\eta}{3}
\left(g_{n+1}\xi_{n+1}-g_{n+2} \xi_{n+2}\right)\right]\, ,
\end{equation} 
together with the condition
\begin{equation}
g_{1}\xi_{1}+g_{2}\xi_{2}+g_{3}\xi_{3}=0\,.
\end{equation}
Moreover, we observe that the asymptotic soliton, whose generic 
expression is 
\begin{equation}\label{soliton}
S(y)=2\eta g\,\exp(i\rho g y)\frac{\gamma^{(+)}\,\gamma^{(-)\ast}}
{|\gamma^{(+)}|^{2}\,
\exp(\eta g\,y)+|\gamma^{(-)}|^2 \exp(-\eta g\,y)}\, ,
\end{equation}
may be  characterized in the usual way by its amplitude $A=E/2$,
 width $\lambda=4/E$ and position $\xi=\log(|\gamma^{(-)} 
/\gamma^{(+)}|)\,
/(\eta g)$ where $E$ is the ``energy"
\begin{equation}\label{solitonenergy}
E=\int^{+\infty}_{-\infty} dy |S(y)|^2\,=\,2|\eta\,g|\,\,.
\end{equation}

Let us now look at one explicit solution. To this aim we choose 
the undressed wave profile
\begin{equation}\label{choice}
f(y)=a\,\exp(-q|y|)\,\,,
\end{equation}
where $a$ and $q$ are arbitrary real parameters ($a=a^*\,,\,q=q^*$)
 and $q$ is positive, $q>0$. The chain of steps we have to make to
 arrive at the asymptotic state expressions (\ref{1asymp}), 
(\ref{2asymp}), (\ref{3asymp})  is the following:
 (i) solving the ZS equations (\ref{ZS}) and obtaining therefore
 also the expressions of $z_1^{(\pm)}(\lambda)$ and 
$z_2^{(\pm)}(\lambda)$, see (\ref{z12}),
 (ii) computing the vector functions $w^{(j)}(y)$, see (\ref{wvector}) 
and (\ref{phimatrix1}), (\ref{phimatrix2}), (\ref{phimatrix3}),
 (iii) computing the constant vector $u^{(j)(\pm)}$, see 
(\ref{u1vector}),
 (\ref{u2vector}), (\ref{u3vector}), and finally (iv) using the
 expressions (\ref{F}) together with (\ref{1soliton}), 
(\ref{2soliton}),
 (\ref{3soliton}). These steps can be made by using the properties of 
Bessel functions $J_{\nu}(z)$ (see for instance \cite{AS}),
and we report here only the resulting expressions with the following 
notation:
\begin{equation}\label{znu}
\zeta=\frac{a}{q}\,\,,\,\,z=\zeta \exp(-q|y|)\,\,,\,\,\nu_{\pm}=\frac{1}{2}\,\pm\,
\frac{i\lambda}{2q}\,\,\,,
\end{equation}
\begin{equation}\label{J}
C_{\pm}=J_{-\nu_{\pm}}^{2}(\zeta)\,-
\, J_{\nu_{\mp}}^{2}(\zeta)\,\,\,.
\end{equation}    
The solution of the ZS equations (\ref{ZS}) we choose is
\begin{subequations}
\begin{equation}\label{ZSsolution1}
\begin{array}{lll}
\phi_1(y,\lambda)&=& \exp\left(-\frac{q}{2}|y|\right)\{H(y)
[C_{-}J_{\nu_{-}}(\zeta)J_{-\nu_{+}}(z)-
C_{+}J_{-\nu_{-}}(\zeta)J_{\nu_{+}}(z)] \\  
&+&H(-y)[C_{-}J_{-\nu_{+}}(\zeta)J_{\nu_{-}}(z)-
C_{+}J_{\nu_{+}}(\zeta)J_{-\nu_{-}}(z)]\}\,\,,
\end{array}
\end{equation}
\begin{equation}\label{ZSsolution2}
\begin{array}{lll}
\phi_2(y,\lambda)&=& \exp\left(-\frac{q}{2}|y|\right)
\{H(y)[C_{-}J_{\nu_{-}}(\zeta)J_{\nu_{-}}(z)+
C_{+}J_{-\nu_{-}}(\zeta)J_{-\nu_{-}}(z)] \\  
&+&H(-y)[C_{-}J_{-\nu_{+}}(\zeta)J_{-\nu_{+}}(z)+
C_{+}J_{\nu_{+}}(\zeta)J_{\nu_{+}}(z)]\}\,\,.
\end{array}
\end{equation}
\end{subequations}
where $H(y)$ is the standard Heaviside step function ($H(y)=1$ if 
$y\ge 0$ 
and $H(y)=0$ if $y<0$). These expressions of $\phi_j(y,\lambda)$ 
 entail that the asymptotic constants (see(\ref{z12})) may 
be chosen as 
\begin{subequations}
\begin{equation}\label{z_1}
z_1^{(\pm)}=\pm \left(\frac{2}{\zeta}\right)^{\nu_{\pm}}
J_{\nu_{\mp}}(\zeta)C_{\mp}\,/\,\Gamma(\nu_{\mp})\,,
\end{equation}
\begin{equation}\label{z_2}
z_2^{(\pm)}=\left(\frac{2}{\zeta}\right)^{\nu_{\mp}}
J_{-\nu_{\mp}}(\zeta)C_{\pm}\,/\,\Gamma(\nu_{\pm})\,,
\end{equation}
\end{subequations}
where $\Gamma$ denotes the Gamma function (see for instance \cite{AS}).
From these expressions all other relevant quantities 
can be computed by simple algebra. For the benefit
 of the reader we report here the standard properties of 
the Bessel and Gamma functions
 which have been used to derive these expressions: 
\begin{equation}\label{bessel}
J_{\nu}(z)=\left(\frac{z}{2}\right)^{\nu}\,\left[\frac{1}{\Gamma(\nu+1)} 
\,+\,O(z^2)\right]\,\,,
\end{equation}
\begin{equation}\label{bessidentity}
z \Gamma(\nu_{+}) \Gamma(\nu_{-}) \left[J_{\nu_{+}}(z) J_{\nu_{-}}(z)
  + J_{-\nu_{+}}(z) J_{-\nu_{-}}(z)\right]=2\,\,,
\end{equation}
the first one of which showing the leading term of $J_{\nu}(z)$
for small $|z|$.

\noindent
It is now  rather straight to play with the various parameters, $a, 
q, \rho, \eta, \gamma_n$ and $c_n$, together with the index 
$j=\,1,\,2,\,3$,
to explore the interaction of the picked 
wave profile $f(y)=a\exp(-q|y|)$, see (\ref{choice}), with solitons in 
a 
variety of different regimes with respect to relative velocities,
energies and soliton positions. Few examples of asymptotic in-- and out--state profiles for such processes are 
graphically displayed in the following figures. 
These examples have been chosen to display processes which may be 
of interest in potential applications of our formulae. 
They show the effect of collision with \emph{one} 
soliton (for $j=1$, see Fig. 1) and with \emph{two} solitons (for $j=2$, see Fig. 2 and Fig. 3) on the 
exponential picked pulse $f(y)$ (\ref{choice}) when such desirable effects as amplitude--amplification
and width--narrowing of the initial in--state profile $f(y)$ occur. 
In all figures the modulus is plotted.

\begin{figure}[h!]
\centering
\fbox{\includegraphics[totalheight=2.5cm]{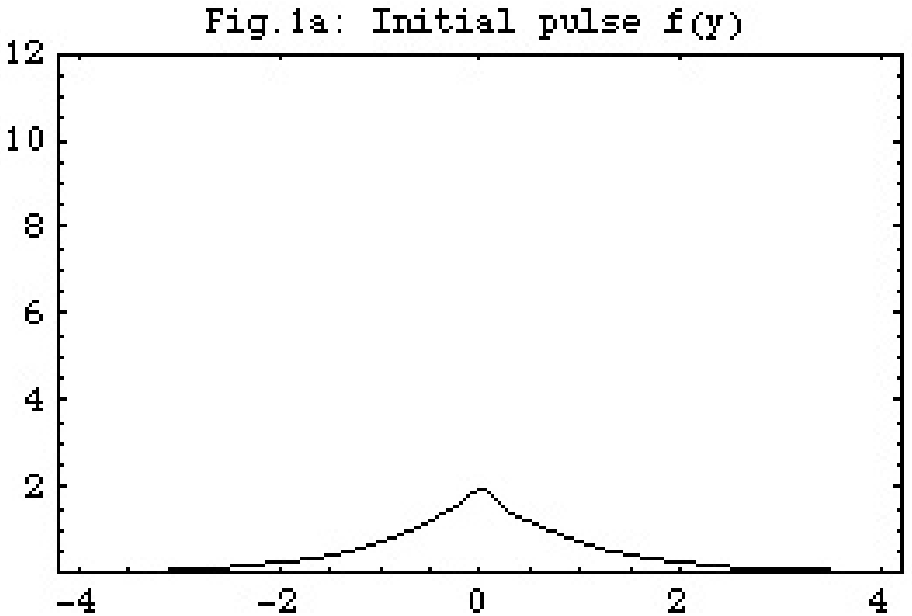}}
\hspace{0.2cm}
\fbox{\includegraphics[totalheight=2.5cm]{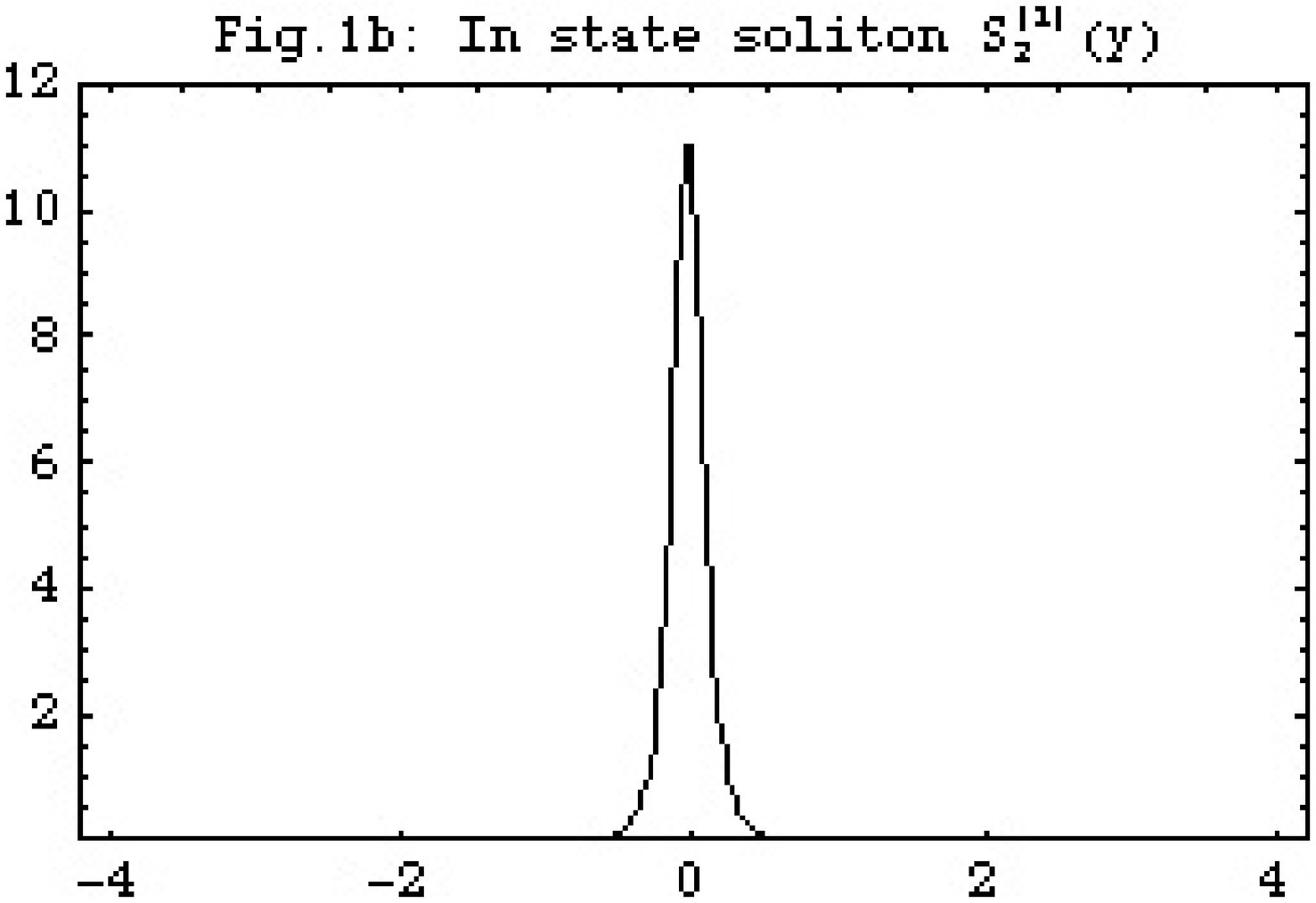}}
\hspace{0.2cm}
\fbox{\includegraphics[totalheight=2.5cm]{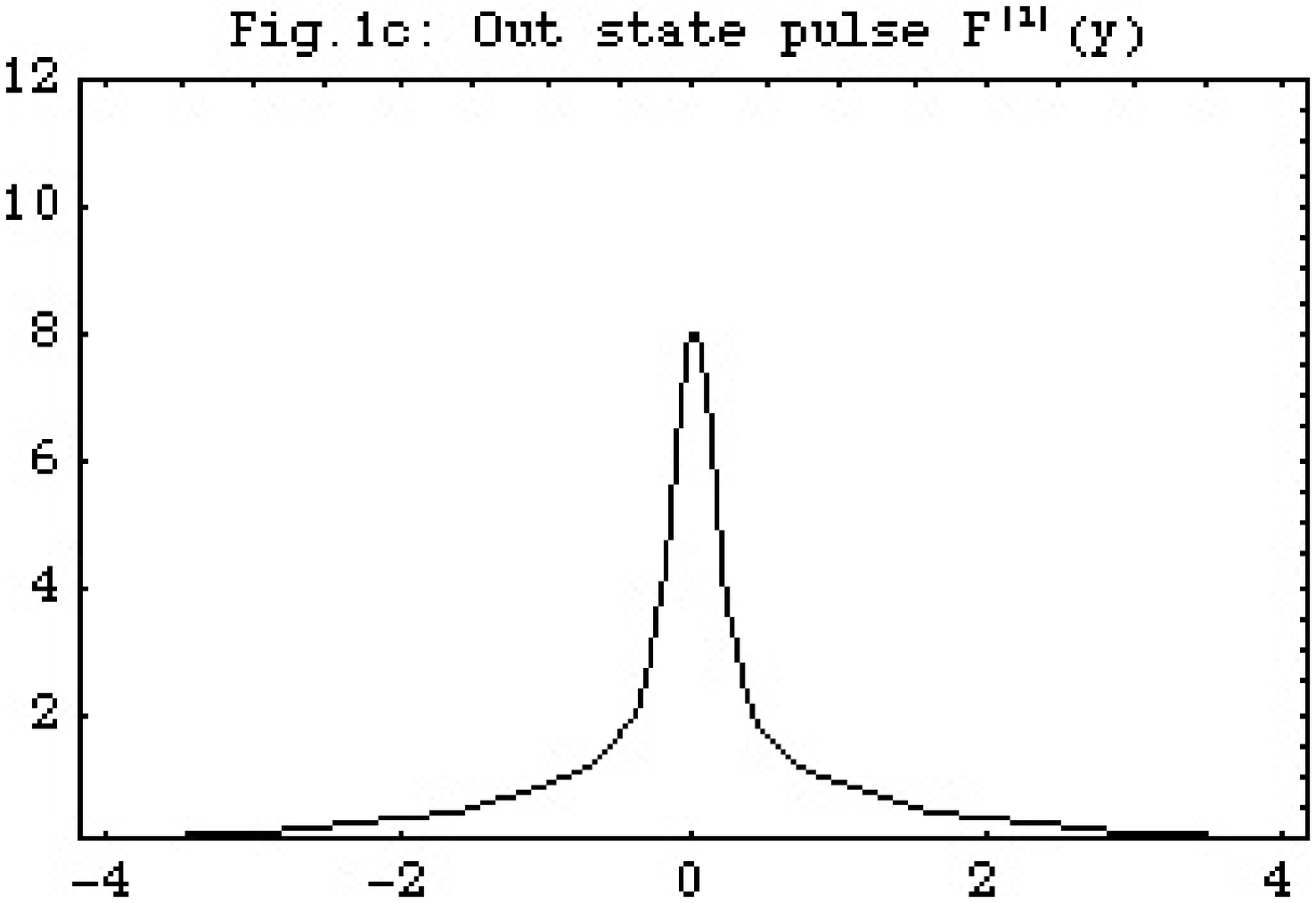}}
\caption{Amplitude amplification.}\label{fig1}
\end{figure}

Fig. 1 refers to the case $j=1$, 
see the asymptotic states (\ref{1asymp}), with the following values of the parameters: 
$c_1=-1$, $\,c_2=0$, $\,c_3=10$, $\,\rho = 0.1$, $\,\eta =-1$, $\,a=2$, 
$\,q=1$, 
$\,\gamma_1 =\gamma_2 =\gamma_3 =1$. 
In particular, Fig. 1a shows the initial pulse $f(y)$, Fig. 1b shows the shape $S^{(1)}_2(y)$, see (\ref{1soliton}), 
of the in--state soliton and Fig. 1c shows the out--state pulse $F^{(1)}(y)$, see (\ref{F}).
Fig. 2 and Fig. 3 refer instead to the case $j=2$, see the asymptotic states (\ref{2asymp}). 
Here two contra--propagating equally shaped solitons with speeds $c_1=-1$ and $c_3=1$ collide 
with the standing ($c_2=0$) initial pulse $f(y)$. 
In  Fig. 2, with the following values of 
the parameters: $\rho = 0.1$, $\,\eta= 1$, $\,a=1$, $\,q=1$, 
$\,\gamma_1 =\gamma_2 =\gamma_3 =1$, 
it is shown the case in which the amplitudes and widths of both the two solitons and the pulse 
$f(y)$ coincide, while in Fig. 3, with the following values of the parameters: 
$\rho = 0.1$, $\,\eta=  2$, $\,a=-1$, $\,q=0.6$, $\,\gamma_1 =\gamma_2 
=\gamma_3 =1$, the amplitude, 
respectively the width, of the colliding solitons is larger, respectively narrower, 
than that of the initial pulse $f(y)$.

\begin{figure}[h!]
\centering
\fbox{\includegraphics[totalheight=2.5cm]{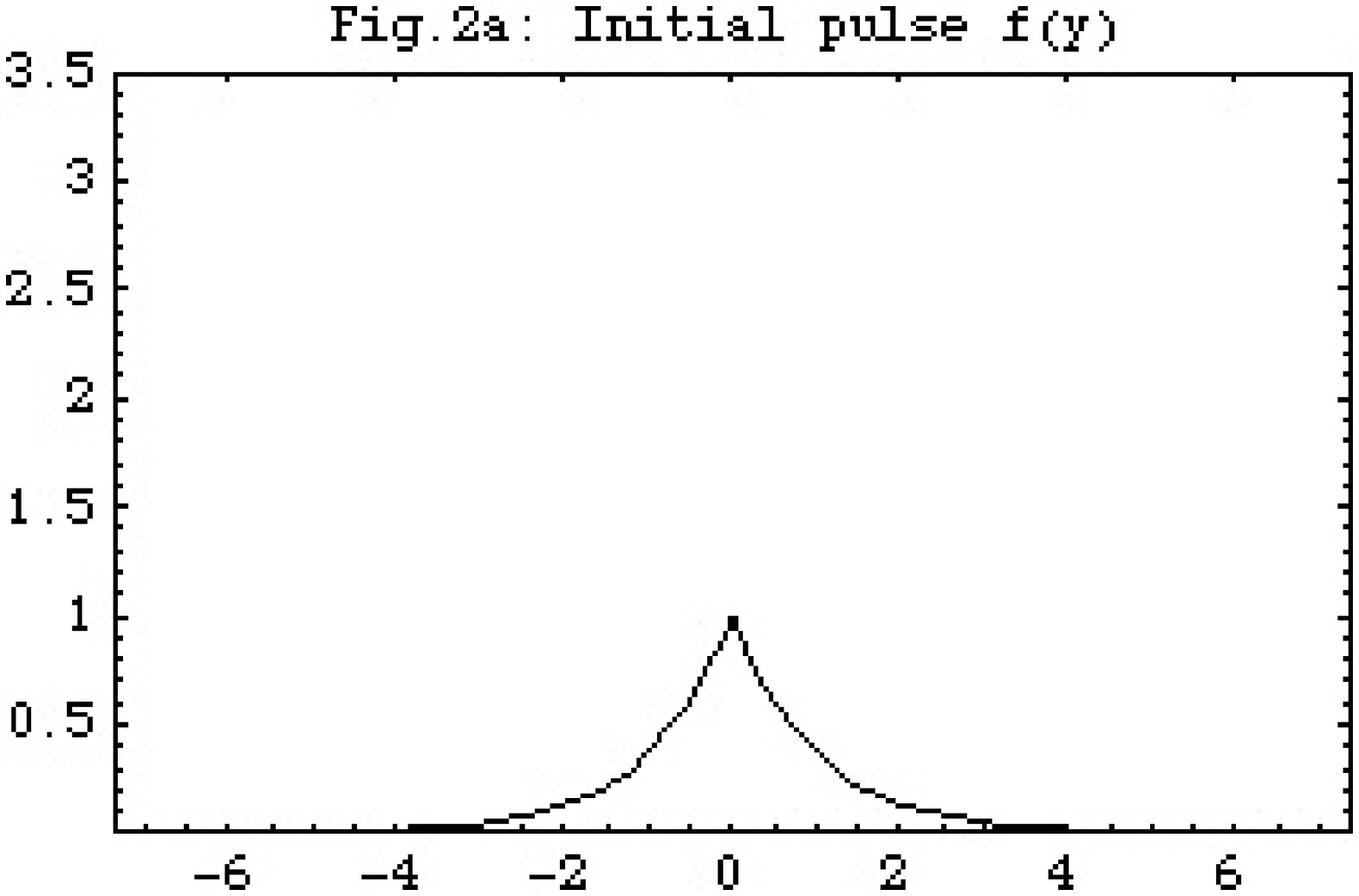}}
\hspace{0.2cm}
\fbox{\includegraphics[totalheight=2.5cm]{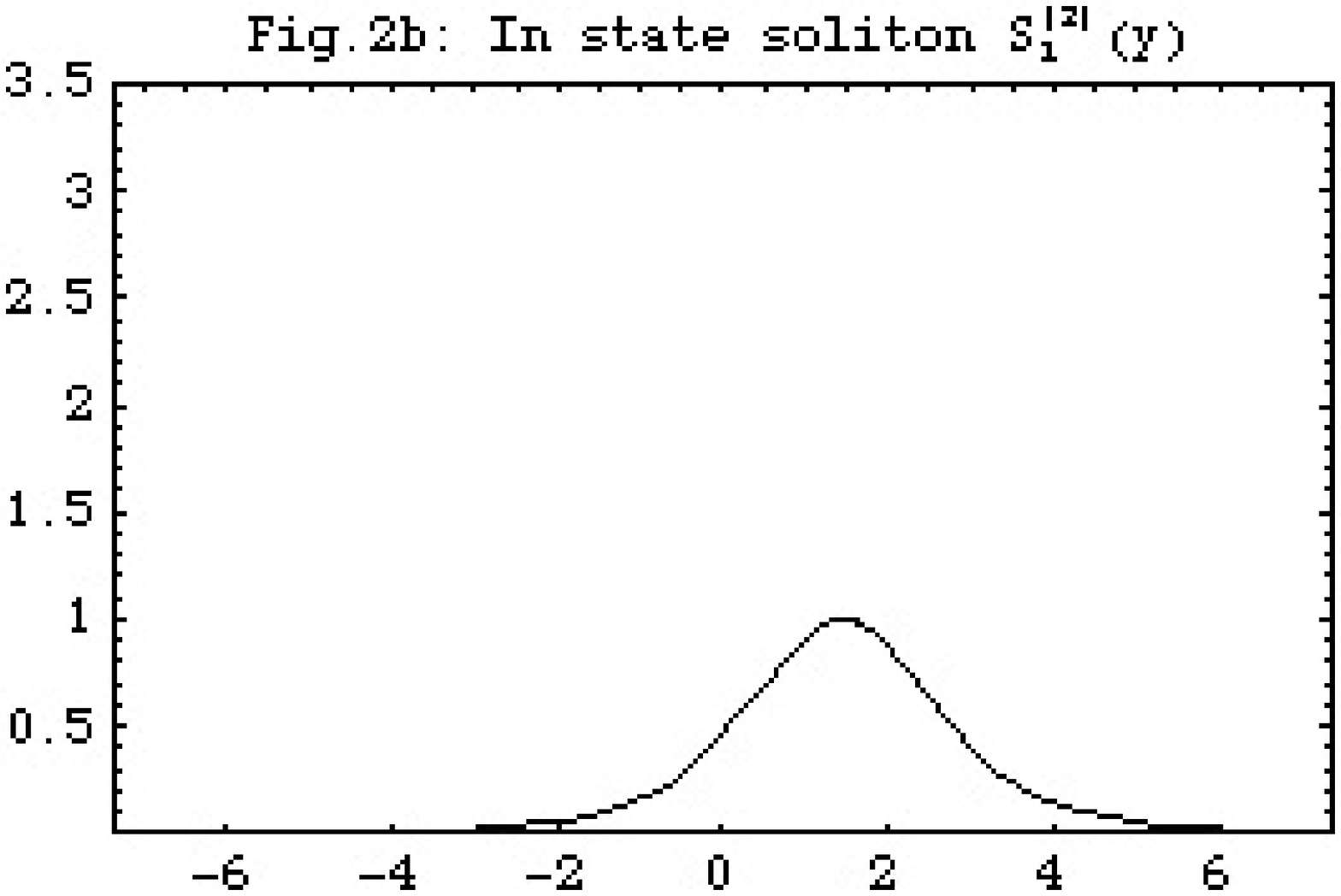}}
\hspace{0.2cm}
\fbox{\includegraphics[totalheight=2.5cm]{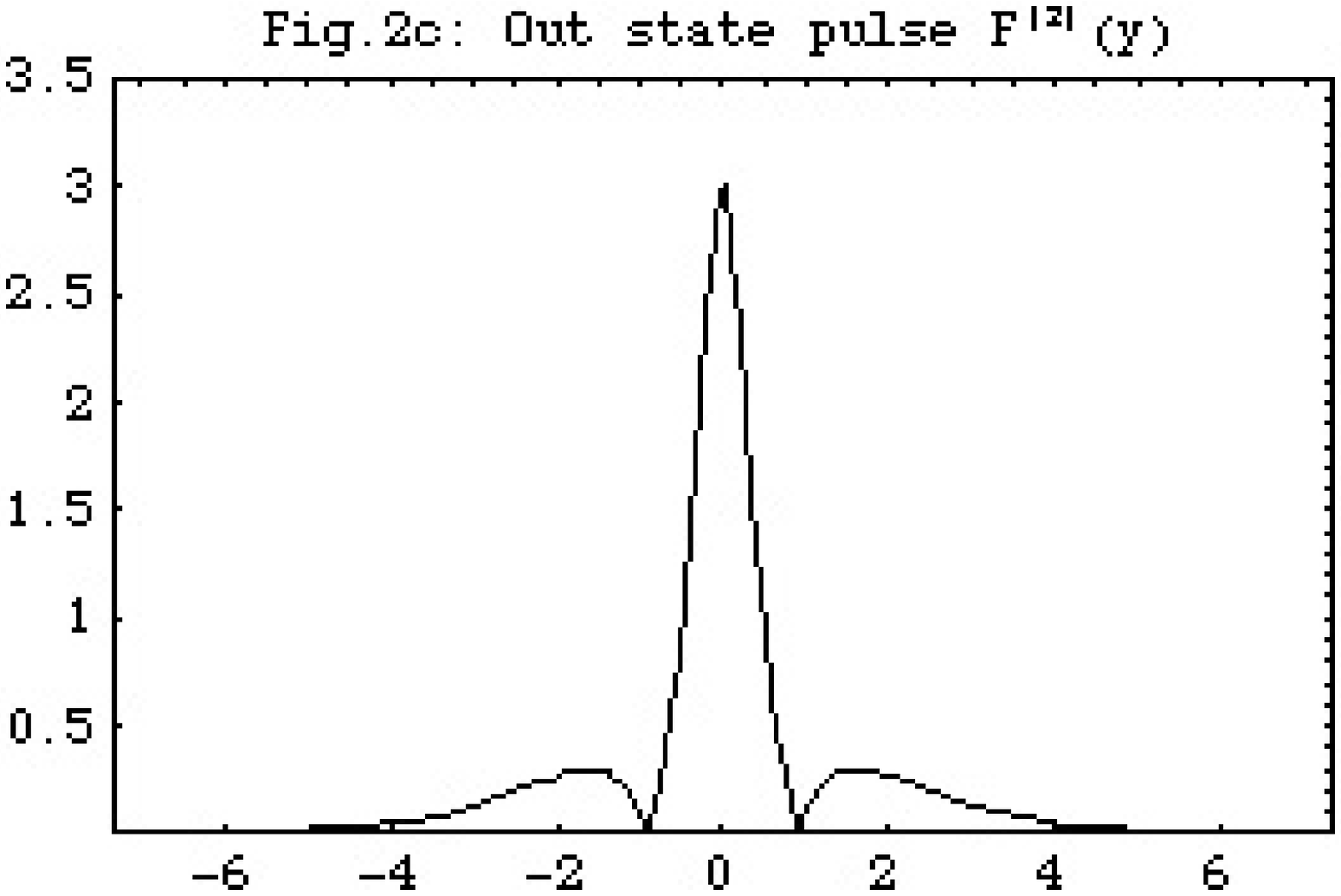}}
\caption{Width--narrowing and amplitude--amplification.}\label{fig2}
\end{figure}

\begin{figure}[h!]
\centering
\fbox{\includegraphics[totalheight=2.5cm]{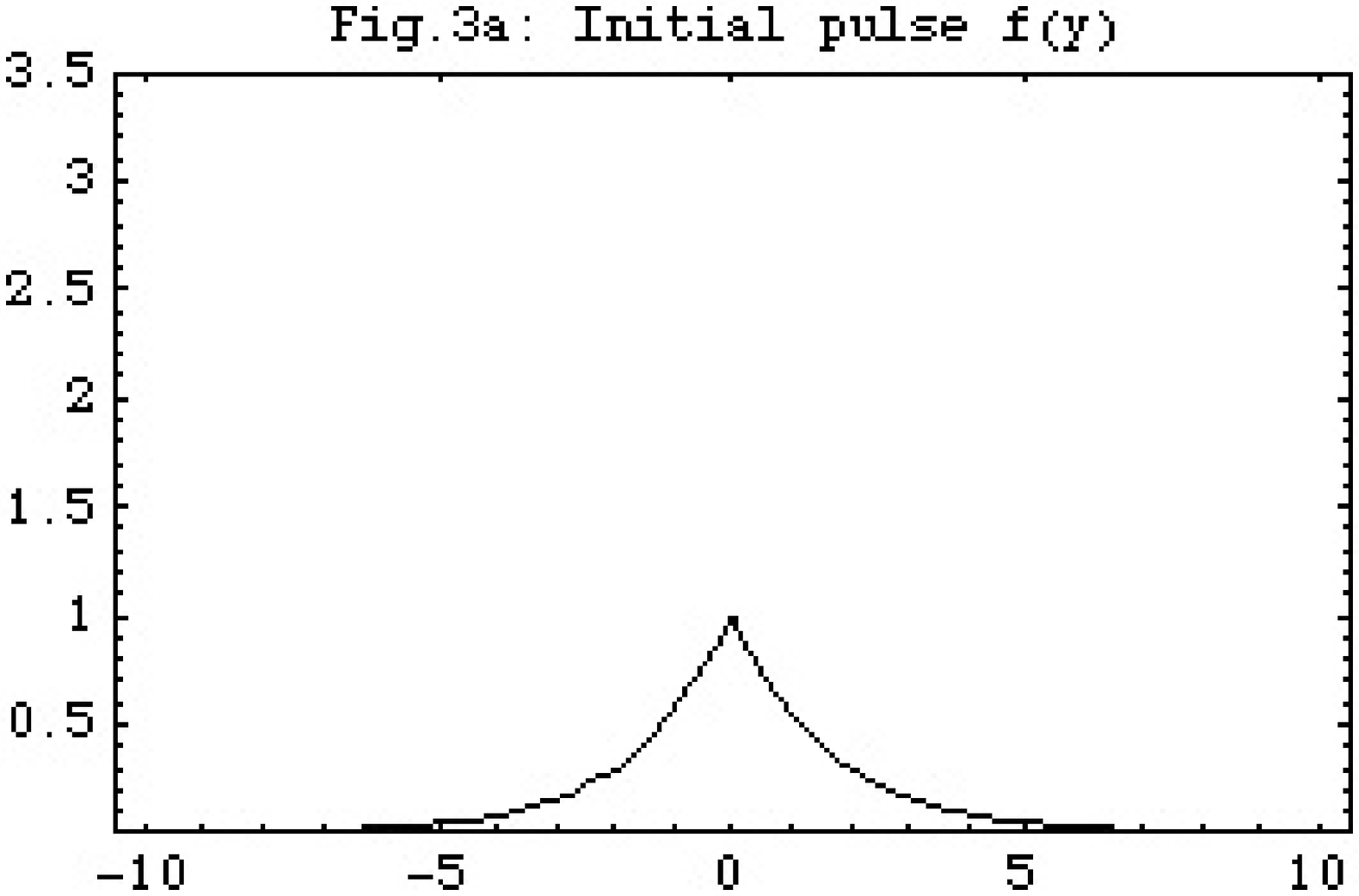}}
\hspace{0.2cm}
\fbox{\includegraphics[totalheight=2.5cm]{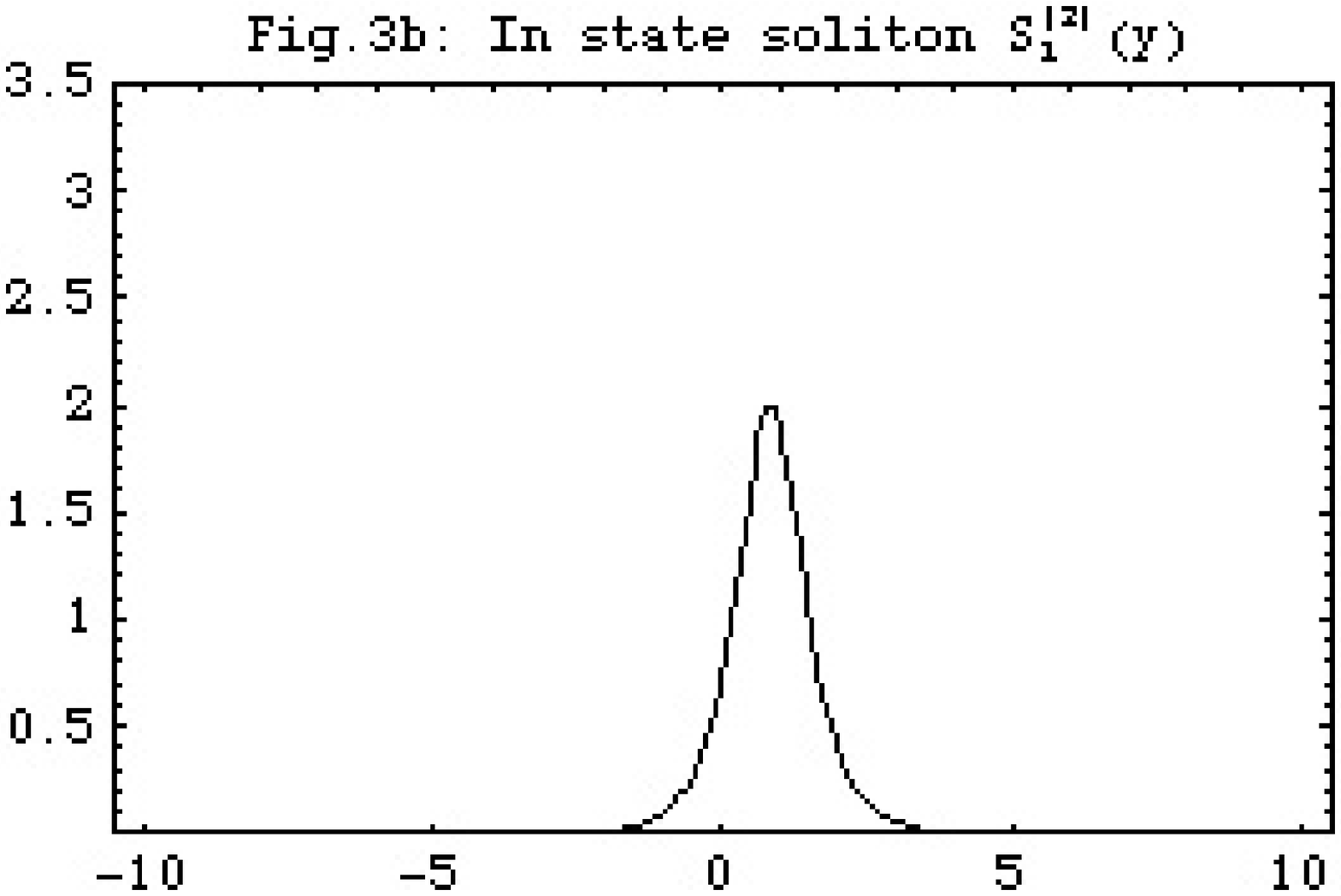}}
\hspace{0.2cm}
\fbox{\includegraphics[totalheight=2.5cm]{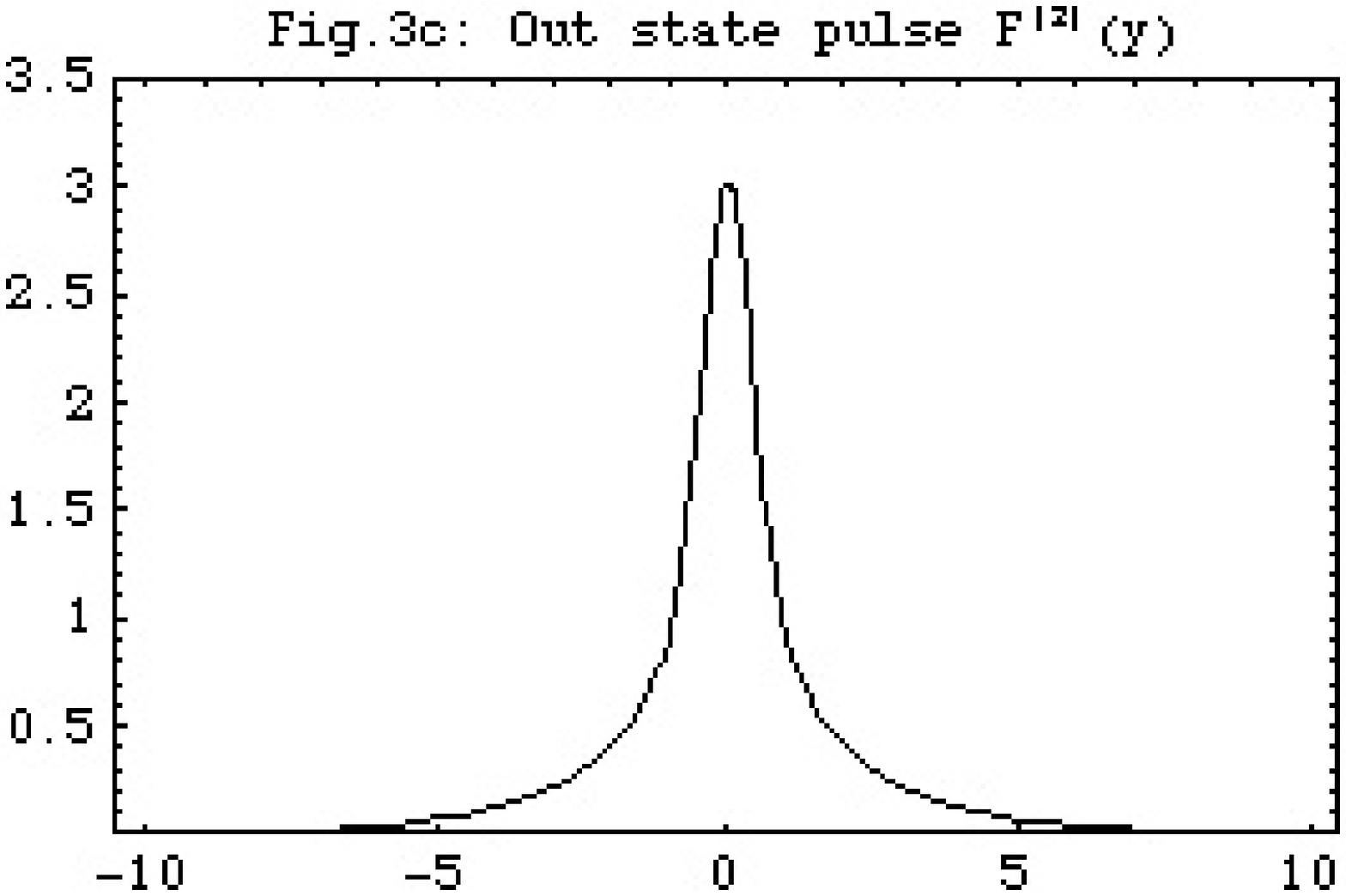}}
\caption{Width--narrowing and amplitude--amplification.}\label{fig3}
\end{figure}

In particular, Fig. 2a and Fig. 3a show the 
initial pulse $f(y)$, Fig. 2b and Fig. 3b show the common shape of the initial solitons 
$S^{(2)}_1(y)$ and  $S^{(2)}_3(y)$ (see (\ref{2asymp}) and (\ref{2soliton})) while Fig. 2c and Fig. 
3c display the out--state profile $F^{(2)}(y)$, see (\ref{F}). Figures in the case $j=3$ are not 
reported since these collisions feature properties similar to those found in the case $j=1$.



We end this section with the observation that our formulae given here are applicable
to several other choices of 
the bare wave $f(y)$ to yield explicit expressions of  solutions of the 3WRI equation, 
provided, of course, that the ZS equations (\ref{ZS}) be exactly solvable. Indeed our choice here,
i.e. $f(y)=a\exp(-q|y|)$, merely indicates how the interested reader may perform simple numerical 
experiments  with explicit analytic expressions. 
\section{Conclusion and remarks}
The system (\ref{3wri}) is a model equation for the resonant interaction 
of three waves. It has been introduced by the method of multiscale
perturbation, as applied to a nonlinear dispersive wave equation, with
the purpose of describing the time evolution of the amplitudes 
$\chi_1, \chi_2, \chi_3$  of three 
plane waves with wave numbers $k_1, k_2, k_3$ and frequencies
$\omega_1, \omega_2, \omega_3$ in the resonance conditions 
$k_1+ k_2+ k_3=0$ and  $\omega_1+ \omega_2+ \omega_3=0$. 
This system is therefore a rather universal model and it finds itself in several
physical applications. Its integrable version, as characterized by the conditions
(\ref{intcond}), has been therefore actively investigated by means of the theory 
of solitons. In particular explicit solutions have been obtained which correspond
to the purely discrete spectrum, i.e. the multi-soliton solutions.
They may be obtained by repeated application of the DDT (see Section 2) 
to the trivial solution $\chi_1= \chi_2= \chi_3=0$. Here we have applied 
the DDT to a seed solution with only two vanishing wave amplitudes while 
the third one is a non vanishing arbitrary function, and we have thereby
constructed a larger class of exact solutions of (\ref{3wri}) outside the class of 
pure discrete spectrum solutions. The feasibility 
of such construction is due to two special properties of the 3WRI system, namely
its lack of dispersion and of self-interaction. We have confined our construction
to the class of localized solutions. However our formulae apply as well to the case 
in which the non vanishing undressed wave $f(y)$, see (\ref{bareUV}), is  
outside the 
class of localized, or finite-energy, wave-profiles. The investigation of 
these ``quasi kink" type solutions is left to future work. One remarkable 
example of such solutions has been recently displayed in \cite{CD3w}. The soliton 
solutions introduced there correspond to those obtained by dressing,
via the method of Section 2, the function $f(y)=a$, $a$ being
an arbitrary complex constant ( this case may be viewed also as the 
$q\rightarrow 0$ (singular) limit of our present choice (\ref{choice})). 
These solutions of the 3WRI equation (\ref{3wri}) describe the 
resonant interaction of two ``bright" solitons with one ``kink" soliton,
a process which features various behaviours, such as those of 
\emph{boomerons} and \emph{trappons}  or the
\emph{creation} and \emph{annihilation} of pairs of bright solitons in the 
background of a kink.  The method of DDT described here yields a rather large class
of new solutions of the 3WRI equation whose generic behavior is of boomeronic 
type, depending on particular choices of the undressed solution. This is indeed the 
first appearance of boomerons as solutions of a model system of  PDEs of such 
wide applicability as the 3WRI equation. In this respect, it should be also pointed out
that  several systems of  coupled Nonlinear Schr\"odinger type equations 
which possess  solutions with similar boomeronic phenomenology have been recently
found and investigated in \cite{CD2004}. 

\section*{Acknowledgments}
Part of this work has been done while attending the \emph{Scientific Gathering on 
Integrable Systems} (October 16--December 11 2004) at the Centro International de Ciencias, 
Cuernavaca, Mexico, and we thank it for partial financial support.   
The work of S L has been partially supported by a grant from \emph{Il Circolo}, Italian Cultural Association, London, UK.


\end{document}